\pdfoutput=1
\documentclass[11pt,a4paper]{article}
\usepackage[T1]{fontenc}
\usepackage[utf8]{inputenc}
\usepackage[margin=25mm]{geometry}
\usepackage{lmodern}
\usepackage{amsmath,amssymb}
\usepackage{graphicx,booktabs,array,float}
\usepackage[normalem]{ulem}
\usepackage{microtype}
\usepackage[numbers,sort&compress]{natbib}
\usepackage[hidelinks]{hyperref}
\newcommand{\ra}[1]{\renewcommand{\arraystretch}{#1}}

\hypersetup{pdftitle={Conformal Holographic Dark Energy and the Neutrino-Mass Tension},
 pdfauthor={Mario A. Rodríguez-Meza, Jorge L. Cervantes-Cota and Tonatiuh Matos}}
\title{Conformal Holographic Dark Energy and the Neutrino-Mass Tension}
\author{Mario A. Rodríguez-Meza\textsuperscript{1},
 Jorge L. Cervantes-Cota\textsuperscript{1}\thanks{Correspondence: \href{mailto:jorge.cervantes@inin.gob.mx}{jorge.cervantes@inin.gob.mx}}
 and Tonatiuh Matos\textsuperscript{2}}
\date{}
\begin{document}
\maketitle
{\small\raggedright\noindent

$^{1}$ \quad Departamento de F\'isica, Instituto Nacional de Investigaciones Nucleares,
Apartado Postal 18-1027, Col. Escand\'on, Ciudad de M\'exico,11801, M\'exico; marioalberto.rodriguez@inin.gob.mx,  jorge.cervantes@inin.gob.mx\\
$^{2}$ \quad Departamento de F\'{\i}sica, Centro de Investigaci\'on y de Estudios Avanzados del Instituto Politécnico Nacional, Av. Instituto Politécnico Nacional 2508, San Pedro Zacatenco, M\'exico 07360, CDMX; tonatiuh.matos@cinvestav.mx
\par\medskip\noindent


\begin{abstract}
Recent DESI results suggest a preference for evolving dark energy (DE) over a cosmological constant, motivating the exploration of alternative models for the background expansion. These data also lead to neutrino-mass constraints close to the lower bounds permitted by oscillation experiments. In this work, we propose and test a conformal holographic dark energy (CHDE) model in which the DE density depends on a power law of the conformal time, characterized by an exponent $n$. The model introduces one additional parameter relative to $\Lambda$CDM and reduces to it in the limit $n=0$.
We confront CHDE with BAO, CMB, and SNe Ia datasets, following the combinations used by DESI, and perform parameter inference in both flat and non-flat cosmologies. We find a consistent preference for $n\simeq -0.25$ to $-0.32$, with uncertainties below $\pm 0.1$, and disfavor the $\Lambda$CDM limit at the $4.4$--$4.5\sigma$ level; however, the model-comparison
statistics provide no consistent preference for
CHDE over $\Lambda$CDM. The resulting neutrino-mass constraints remain compatible with terrestrial limits and yield posterior distributions that peak at positive values, avoiding the negative-mass preference found in $\Lambda$CDM. The CHDE model is therefore consistent with current observational constraints and provides a viable alternative to constant dark energy. 
\end{abstract}


\section{Introduction}
The standard model of cosmology is based on a content of baryons, photons, neutrinos, cold dark
matter, and dark energy (DE). Although we do not know the origin of cold dark matter, we have characterized
its gravitational effects at different astrophysical and cosmological scales \cite{Bertone:2016nfn}, and it seems consistent with
its cold origin, but interesting alternatives have been discussed in the literature such as warm and fuzzy
dark matter \cite{MatosUrenaLopez2001,Hui:2016ltb}. On the other hand, we know less about DE, but it is clear that a cosmological constant is the simplest way to account for the cosmic acceleration. For many years, the cosmological constant plus cold dark matter ($\Lambda$CDM) model has been
remarkably successful in explaining a wide range of cosmological observations, including Type Ia supernovae (SNe Ia) distance
measurements, cosmic chronometers, the Cosmic Microwave Background (CMB), and large-scale structure clustering
statistics. However, the cosmological constant has recently been called into question by the joint analysis of
Baryon Acoustic Oscillations (BAO) performed by the Dark Energy Spectroscopic Instrument (DESI) collaboration Year 1
\cite{DESI:2024mwx,DESI:2024aqx,DESI:2024kob} and Year 2 \cite{DESI:2025zgx,DESI:2025kuo}, DES final BAO
\cite{DES:2025bxy}, and supernovae results: Pantheon+ \cite{Scolnic:2021amr,Brout:2022vxf}, Union3
\cite{Rubin:2023jdq}, and the Dark Energy Survey (DES) Year 5 \cite{DES:2024jxu}. Their analysis suggests that the DE  responsible for
the current epoch of cosmic acceleration is decreasing at present. In any case, the results provide evidence
in favor of a dynamical form of dark energy rather than a strictly constant one.  Because the origin of this
behavior remains unclear, numerous alternative explanations have been proposed. These can be broadly classified
into fundamental theoretical models and parameterized  phenomenological models.

As far as we know, cosmological data does not reveal DE clustering properties, but it mainly manifests its
effect in the  background dynamics (apart from the the integrated Sachs–Wolfe (ISW) effect). Therefore, when thinking of passing from a
cosmological constant to a function, it seems appealing to associate a cosmological function with no clustering
properties, or, if any, at very large distances such as horizon scales, and that is why a holographic Universe
is accommodated in a natural way. Holographic dark energy (HDE) has been a popular model for years, and
different realizations of it have been put forward \cite{Nojiri:2005pu,Wang:2016och,Nojiri:2021iko}. Here we discuss  a holographic model
that depends on the conformal time. However, our model is different from the well-known agegraphic models
\cite{Wang:2016och,Wei:2007ty}, in which matter conservation is defined apart from the Einstein equations,
which are assumed to be valid. In our case, the cosmological DE function will be explicitly written in the Einstein
equations, so energy conservation demands an interaction between our HDE model and dark matter in particular.
Our model has the mathematical properties of a vacuum DE model as the one presented in refs. \cite{Wands:2012vg,De-Santiago:2012xpd}.

On the other hand, recent cosmological measurements of the
neutrino mass by the DESI collaboration \cite{DESI:2025zgx}
suggest extremely small values that may conflict with limits
established by terrestrial neutrino oscillation experiments, see also \cite{CosmoVerseNetwork:2025alb}. Some
DESI data combinations even seem to disfavor both the normal and
inverted mass hierarchies, and the preferred values in the
statistical analyses tend to lie in nonphysical, negative regions
\cite{DESI:2025ejh}. These tensions motivate the exploration of
alternative cosmological scenarios. One proposal is to allow
spatial curvature to vary, which weakens the constraints and
shifts the inferred values toward more physically plausible ones
\cite{Chen:2025mlf}. However, even in this case, the most likely
value remains close to zero, indicating that a full resolution of
the issue may require analyses in which effectively negative
neutrino masses are allowed. Indeed, this latter analysis suggests
that curvature alone does not provide a solution to the tension
\cite{Pulido-Hernandez:2026hcs}. In this context, DE models
should be examined to assess whether they can resolve, or at least
alleviate, the tension in the inferred neutrino mass. For
instance, the \texorpdfstring{$w_{0}w_{a}$CDM}{w0waCDM} model
\cite{Chevallier:2000qy,Linder:2002et,Pulido-Hernandez:2026hcs}
provides partial relief, although it does not fully eliminate the
problem once all data sets are included. Other DE models appear to resolve the issue for particular choices of the equation of state that cross the phantom divide \texorpdfstring{($w=-1$)}{} \cite{Yang:2026yaq}, whereas our model keeps \texorpdfstring{$w =-1$}{}, with the time dependence entering through the interaction term. In this work, we investigate this question within the framework of our proposed model.

The aim of this paper is to develop and observationally test
a conformal holographic dark energy (CHDE) model as an alternative
to $\Lambda$CDM, and to assess whether it can alleviate the tension
between cosmological neutrino-mass constraints and the lower bounds
inferred from terrestrial oscillation experiments. In this model, the
dark-energy density follows a power law in conformal time and
exchanges energy with dark matter while retaining an equation of
state $w=-1$. We implement the model in the Cosmic Linear Anisotropy Solving System (\texttt{class}) and perform
Bayesian parameter inference using
Markov chain Monte Carlo (MCMC) sampling
with \texttt{Cobaya}, employing combinations of DESI Data Release 2 (DR2) BAO,
Planck CMB, Atacama Cosmology Telescope (ACT) lensing, and Type Ia supernova data. We consider both
spatially flat and non-flat cosmologies to constrain the model
parameters, examine the inferred sum of neutrino masses, and
compare the model's statistical performance with that of $\Lambda$CDM.

This work is organized as follows: In section \ref{sect:model} we present the CHDE model and derive the relevant equations. In section \ref{sect:model_testing} we employ different datasets and free parameters to extract the physical information of the model. Finally, in section \ref{conclu} we review our results and conclude. One appendix is provided that presents our theoretical implementation in \texttt{class}.

\section{The CHDE model}  \label{sect:model}
We start with the Einstein equations with a holographic term given by \cite{Matos:2023qwx}:
\begin{equation}\label{eq:Einstein}
R^{\mu\nu}-\frac{1}{2} R g^{\mu\nu}+ \mathcal{M} g^{\mu\nu}=\kappa^2 T^{\mu\nu},
\end{equation}
where $\kappa^2=8\pi G/c^4$; $G$ is Newton's gravitational constant, $T^{\mu\nu}$ contains the standard matter pieces of the standard cosmological model (baryons, cold dark matter (DM), and relativistic particles), and $\mathcal{M}$ is a holographic function that will depend on the horizon distance and will play the role of a cosmological, dark energy function. One may postulate different holographic functions, see e.g. \cite{Wang:2016och}, which typically depend on the physical horizon. But one can also consider the comoving horizon. In ref.
\cite{Matos:2021jef}, the authors opted for this option taking $\mathcal{M} \sim \frac{1}{\lambda^2}$, where $\lambda$ is the cosmological length scale given by $\lambda = \frac{c}{H_0} R_H$, where $R_H =H_0 \int_{0}^{t} dt^{'}/a(t^{'})$ is the Friedmann–Robertson–Walker (FRW) comoving horizon, where $a(t)$  is the scale factor, $t$ the cosmic time and $H_0$ the Hubble constant. This model was motivated by its connection to de Proca gravity and the interpretation that it represents a gravitational wave background \cite{Matos:2021jef}.
The interpretation of the previous model in terms of a gravitational-wave background motivates exploring whether a dependence of the wave amplitude on the cosmological length scale $\lambda$ could modify the inverse-square
scaling. Here, we adopt a phenomenological power-law generalization,
rather than a first-principles derivation from that interpretation. Since $R_H=H_0 \eta$, it is convenient to express the model in terms of conformal time, defined by $d\eta=dt/a(t)$. Our conformal holographic dark energy (CHDE) model is
\begin{equation}\label{eq:HF}
\mathcal{M} = const. \,  \eta^n ,
\end{equation}
where the exponent $n$ is a constant and the $const$ is conveniently defined as $const= Q \,  H_0^{n+2} $, with $Q$ a dimensionless constant of order one related to the cosmological function evaluated today and $H_0^{-1}$ is of the order of the horizon at present that gives the proper units to $\mathcal{M}=Q H_0^2(H_0\eta)^n$, being $H_0\eta$ dimensionless. Equation (\ref{eq:HF}) is a phenomenological ansatz motivated by the conformal-horizon
dependence of the previously studied model. The power-law form provides
a simple one-parameter extension of its inverse-square scaling:
$n=-2$ recovers the previous model, whereas $n=0$ gives a constant
dark-energy density and recovers $\Lambda$CDM. Allowing $n$ to vary
therefore enables an observational test of departures from these two
limits. The exponent $n$ is constrained by cosmological observations,
while $Q$ is determined through the present-day closure
relation, as described below.

Note that it is convenient to define $\rho_{\mathcal{M}} = \mathcal{M}/8 \pi G$ and therefore
\begin{equation}\label{eq:HF2}
\Omega_{\mathcal{M}} = \frac{\rho_{\mathcal{M}}}{\rho_{c0}} = \Omega_{\mathcal{M}}^{(0)} \left(\frac{\eta}{\eta_0}\right)^n \, ,
\end{equation}
where $\Omega_{\mathcal{M}}^{(0)} \equiv \frac{1}{3}Q (H_{0} \eta_0)^{n}$ and $\rho_{c0}$ the standard critical density. Eq. (\ref{eq:HF2}) is our cosmological function, in which the cosmological scale, $\eta_0 \sim H_{0}^{-1}$ is the comoving scale evaluated  at present times; in principle, it can be thought of as a fundamental scale responsible for DE.  Assuming a FRW background geometry, the gravitational equations (\ref{eq:Einstein}) imply

\begin{equation}\label{eq:Hubble1}
\frac{{\cal{H}}^{2}}{H_{0}^{2}} =  \frac{\Omega_{b}^{(0)}}{a} + \frac{\Omega_{\gamma}^{(0)}}{a^2} + \Omega_{k}^{(0)} +
\Omega_{dm}^{(0)}  \, \frac{\rho_{dm}}{\rho_{dm}^{(0)}} \, a^{2} +  \Omega_{\mathcal{M}}^{(0)} a^2 \, \left(\frac{\eta}{\eta_0}\right)^{n},
\end{equation}
where ${\cal{H}} \equiv \frac{a^{'}}{a} = \frac{1}{a} \frac{d a}{d \eta}$; we will also consider the curvature that can be added to this equation.
For a fixed value of $n$, the CHDE model has the same number of free parameters as $\Lambda$CDM. In the fits presented here, however, $n$ is varied freely, introducing one additional free parameter relative to $\Lambda$CDM.
Note that one can recover the $\Lambda$CDM model for $n=0$;  $n=-2$ was used in previous works by one of the authors \cite{Matos:2021jef,Matos:2023qwx}. On the other hand, if there is no interaction between DM and CHDE ($Q =0$), $\rho_{dm} \sim 1/a^3$, one recovers the Einstein de Sitter Universe (with no radiation and curvature).
But, in general, matter conservation demands

\begin{equation}\label{eq:conserva}
\rho_{dm}^{'} + 3  {\cal{H}} \rho_{dm} = - \rho_{\mathcal{M}}^{'} = - n \frac{\rho_{c}^{(0)}}{\eta_0} \Omega_{\mathcal{M}}^{(0)} \, \left(\frac{\eta}{\eta_0}\right)^{n-1} \, .
\end{equation}
where $\rho_{c}^{(0)} \equiv 3 H_{0}^{2}/8 \pi G$. This equation specifies the interaction of DM with CHDE, in which one can directly observe that DM is created if $n<0$, and it is sourced by a time decaying function for $n<1$. The CHDE is a geometric term, but in analogy to the $\Lambda$CDM model, it can be thought of as a perfect fluid, so that the r.h.s. of eq. (\ref{eq:conserva}) is equal to $-[\rho_{\mathcal{M}}^{'} + 3  {\cal{H}} (1+{\rm w}_{\mathcal{M}}) \rho_{\mathcal{M}}]$, therefore, this equation shows that ${\rm w}_{\mathcal{M}} =-1$. This can also be seen from eq. (\ref{eq:Einstein}): the relationship between $\rho_{\mathcal{M}}$ and $p_{\mathcal{M}}$ is given similar to the $\Lambda$CDM model, $T_{\mathcal{M}}^{\mu \nu} = - \mathcal{M} g^{\mu \nu}/\kappa^2$, from which one can read off the equation of state (EoS) for $\mathcal{M}$. This derivation is similar to that presented in references \cite{Wands:2012vg,De-Santiago:2012xpd}.

Eq. (\ref{eq:Hubble1}) evaluated at present, $\eta = \eta_0$, imposes a constraint on the DE,
\begin{equation}\label{eq:constraint}
1=\Omega_{b}^{(0)} + \Omega_{\gamma}^{(0)} + \Omega_{k}^{(0) +} \Omega_{dm}^{(0)} + \Omega_{\mathcal{M}}^{(0)} \, .
\end{equation}
The new complexity is that this constraint demands computing $H_{0} \eta_0$, so we implemented a check of this constraint  at every time step.

Since we aim to solve the system of equations (\ref{eq:Hubble1}) and (\ref{eq:conserva}), with the constraint equation \ref{eq:constraint}, using \texttt{class} \cite{Blas:2011rf}, we write in appendix \ref{app:B} the corresponding \texttt{class} equations, in which we make use of the conformal and cosmic times, as well as the number of e-folds parameter $N (\equiv \text{ln}\,  a)$. The last term of eq. (\ref{eq:Hubble1}) (or \texttt{class} version \ref{eqn:H2}) decreases in the past for $n>0$, but diverges for $n<0$, however in a matter- or radiation-dominated universe, the energy density diverges too as  $\rho_{dm} \sim 1/\eta^{6}$ and $\rho_{\gamma} \sim 1/\eta^{4}$, respectively. Thus, the final result for $n$ should not diverge faster than $n=-4$ to maintain our CHDE model subdominant in these Universe's eras, only becoming dominant at later times, $z < 1$.
In figure \ref{fig:densities}, we show the densities' evolution, demonstrating the mentioned behavior, where we observe the decrease of dark matter, steeper than baryons, and the rise of dark energy as time evolves.

\begin{figure*}[h]
	\centering
	\includegraphics[width=0.80\textwidth]{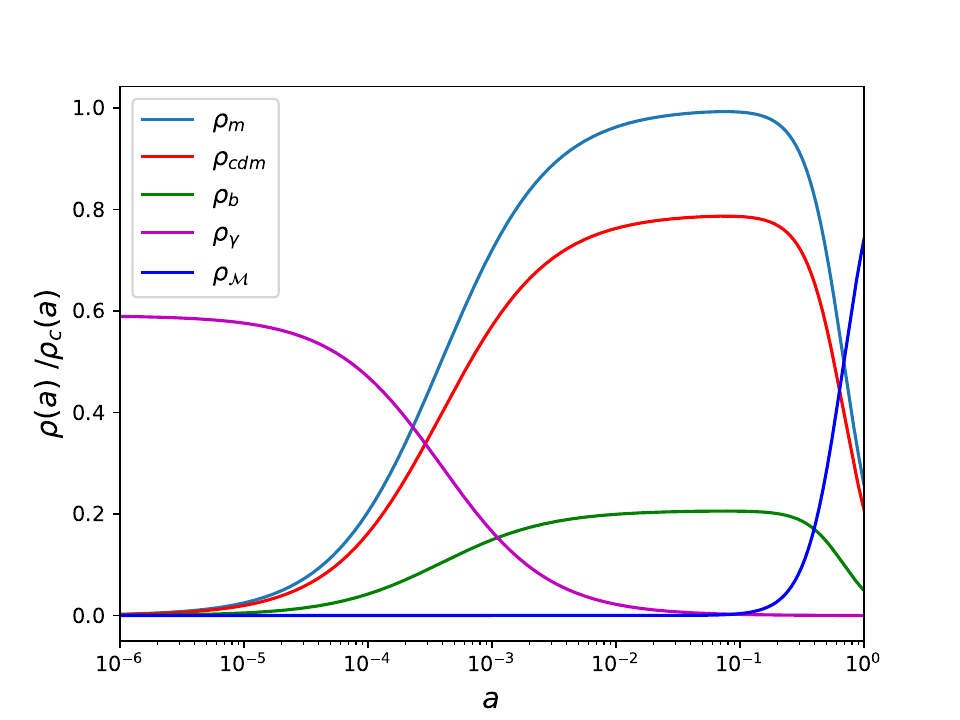}
\caption{Density parameters evolution as a function of the scale factor.} \label{fig:densities}
\end{figure*}

We now clarify the perturbative prescription adopted in this work. The standard perturbation equations are retained for photons,
baryons, and neutrinos, while the dark-matter continuity
equation includes the interaction term derived below.
The metric perturbations evolve consistently with these
components and the modified background. The background continuity equation alone does not uniquely determine the perturbations:
a covariant prescription for the energy transfer is also required
\cite{Wands:2012vg}. We adopt the covariant energy-transfer prescription
\begin{equation}
    {\bf Q}_\mu = {\bf Q} \,u^{({dm})}_\mu ,
    \label{eq:geodesic_transfer}
\end{equation}
i.e. the energy flow is parallel to the dark-matter four-velocity. This
choice implies that there is no momentum transfer in the dark-matter rest
frame. In our case,
\begin{equation}
    T^{\mu}_{\mathcal{M}\, \nu}
    =
    -\rho_{\mathcal{M}} \,\delta^{\mu}{}_{\nu}.
\end{equation}
In particular, the gauge-invariant CHDE density perturbation on
hypersurfaces comoving (com) with dark matter satisfies
\begin{equation}
    \delta\rho_{\mathcal{M},{\rm com}}
    =
    \delta\rho_\mathcal{M} + \rho_\mathcal{M}'\theta_{dm}
    =
    0 .
\end{equation}
Hence, in the dark-matter-comoving synchronous gauge,
\begin{equation}
    \theta_{dm}=0,
    \qquad
    \delta\rho_\mathcal{M}=0.
\end{equation}
The CHDE component therefore introduces no independent propagating scalar
degree of freedom in this prescription. Nevertheless, the interaction
modifies the evolution of the fractional dark-matter density contrast through
the time dependence of the background dark-matter density.
For pressureless dark matter, the perturbed conservation equation of
Ref.~\cite{Wands:2012vg} reduces, in the dark-matter-comoving synchronous
gauge, to
\begin{equation}
    \delta\rho_{dm}'
    +3{\cal H}\,\delta\rho_{dm}
    +\frac{1}{2}\rho_{dm}h'
    =
    0,
    \label{eq:delta_rho_dm}
\end{equation}
where \(h\) is the usual scalar metric perturbation in synchronous gauge.
Introducing the fractional contrast
\begin{equation}
    \delta_{dm}
    \equiv
    \frac{\delta\rho_{dm}}{\rho_{dm}},
\end{equation}
and using Eq.~\eqref{eq:conserva}, one obtains
\begin{equation}
    \delta_{dm}'
    =
    -\frac{h'}{2}
    +
    \frac{\rho_\mathcal{M}'}{\rho_{dm}}\,
    \delta_{dm}.
    \label{eq:delta_dm_chde_general}
\end{equation}
For the CHDE density this becomes
\begin{equation}
    \delta_{dm}'
    =
    -\frac{h'}{2}
    +
    \frac{n\rho_\mathcal{M}}{\eta\,\rho_{dm}}\,
    \delta_{dm}
    .
    \label{eq:delta_dm_chde}
\end{equation}
Thus, although the CHDE density perturbation itself vanishes in the
dark-matter-comoving synchronous gauge, the vacuum--dark-matter energy
exchange changes the growth of the dark-matter density contrast. All other
standard species are evolved with their usual perturbation equations. In the
limit \(n=0\), one has \(\rho_\mathcal{M}'=0\), and
Eq.~\eqref{eq:delta_dm_chde} reduces to the standard CDM result
\(\delta_{dm}'=-h'/2\). In appendix \ref{app:B} we include the perturbed equation \ref{eqn:classB5}.

\section{Cosmological data and model testing} \label{sect:model_testing}

\subsection{Cosmological data}

We implement the CHDE model in \texttt{class} \cite{Diego_Blas_2011} and perform Bayesian
parameter inference through Markov chain Monte Carlo sampling with
\texttt{Cobaya} \cite{Torrado:2020dgo}. The observational datasets and their implementation
are summarized below.

\begin{itemize}

\item \textbf{DESI DR2 BAO.}
We use the compressed baryon acoustic oscillation measurements
from DESI Data Release~2 \cite{DESI:2025zgx}, for short, BAO. These constrain combinations
of the transverse comoving distance $D_M(z)$, the radial distance
$D_H(z)=c/H(z)$, and the volume-averaged distance
$D_V(z)=[zD_M^2(z)D_H(z)]^{1/3}$, expressed relative to the
sound horizon at the baryon drag epoch, $r_d$.
We employ  the public DESI DR2 BAO likelihood implemented in
\texttt{Cobaya}, with the released compressed distance measurements
and their covariance matrix, including statistical and systematic
uncertainties \cite{DESI:2025zgx}. The likelihood is taken to be
multivariate Gaussian,
\begin{equation}
-2\ln\mathcal{L}_{\mathrm{BAO}}
=
\Delta\mathbf{d}^{T}
\mathbf{C}_{\mathrm{BAO}}^{-1}
\Delta\mathbf{d}
+\mathrm{const.},
\end{equation}
where $\Delta\mathbf{d}=\mathbf{d}_{\mathrm{th}}
-\mathbf{d}_{\mathrm{obs}}$ and the theoretical distance ratios
are computed with our modified \texttt{CLASS} code.

\item \textbf{Planck CMB.}
 We employ the multipole \texttt{CamSpec} likelihood (version~12.6), based on the Planck PR4/NPIPE maps, using
the combined $TT$, $TE$, and $EE$ spectra \cite{Efstathiou:2019mdh,Rosenberg:2022sdy}.
The theoretical spectra are computed with the modified
\texttt{CLASS} code, including the CHDE background evolution
and the perturbation prescription described in Section \ref{sect:model}.

\item \textbf{ACT lensing.}
We include the ACT Data Release 6 (DR6) CMB lensing reconstruction \cite{ACT:2023kun,ACT:2023dou,ACT:2023ubw}. This dataset constrains the angular power
spectrum of the CMB lensing potential, $C_L^{\phi\phi}$, providing
information on the intervening matter distribution and its growth.
We use the public ACT DR6 CMB lensing likelihood
, adopting the
ACT-only baseline reconstruction, which combines temperature
and polarization information. The likelihood assumes Gaussian
errors for ten lensing power-spectrum bandpowers over
$40<L<763$ and uses the released simulation-based covariance
matrix.

\item \textbf{Pantheon+.}
This compilation combines standardized Type~Ia supernova
light curves from several surveys \cite{Scolnic:2021amr,Brout:2022vxf}.
We use the Pantheon+ compilation implemented through the standard
\texttt{sn.pantheonplus} likelihood in \texttt{Cobaya}.
We select objects with Hubble-diagram redshifts
$z_{\mathrm{HD}}>0.01$ and adopt the released covariance
matrix including statistical and systematic uncertainties,
restricted to the selected sample. The likelihood is Gaussian
in the corrected apparent-magnitude residuals, with the
absolute-magnitude normalization analytically marginalized.
We do not include the SH0ES Cepheid calibration.

\item \textbf{Union3.}
Union3 combines Type~Ia supernova observations analyzed with the
UNITY1.5 Bayesian hierarchical framework \cite{Rubin:2023jdq}.
We use the compressed Union3 distance measurements derived
with the UNITY1.5 Bayesian hierarchical framework. We employ the standard \texttt{sn.union3}
likelihood in \texttt{Cobaya}, which compares the released
distance--redshift relation with the luminosity distances
predicted by the CHDE model. The likelihood adopts a
multivariate Gaussian approximation with the supplied
covariance matrix, incorporating statistical and systematic
uncertainties and correlations between the compressed
distance measurements. The overall distance normalization
is analytically marginalized, so no external absolute-magnitude
calibration is imposed.

\item \textbf{DES Year~5.}
We use the DES Year~5 supernova sample \cite{DES:2024jxu},
comprising 1635 photometrically classified supernovae in
the range $0.10<z<1.13$, together with 194 external
low-redshift Type~Ia supernovae at $0.025<z<0.10$.
We adopt the released cosmological sample without
additional selection cuts and employ the standard
\texttt{sn.desy5} likelihood in \texttt{Cobaya}.
The likelihood is Gaussian in the distance-modulus
residuals relative to the CHDE prediction. Its covariance
combines the released systematic covariance matrix with
the diagonal statistical variances supplied with the
distance measurements. The overall magnitude normalization
is analytically marginalized, without imposing an external
absolute-magnitude calibration.

\end{itemize}

For the supernova datasets, the predicted distance modulus is
\begin{equation}
    \mu_{\mathrm{th}}(z)
    =5\log_{10}\left[\frac{D_L(z)}{\mathrm{Mpc}}\right]+25,
\end{equation}
where $D_L(z)$ is the luminosity distance in the CHDE cosmology.
Pantheon+, Union3, and DES Year~5 are considered as alternative
supernova compilations, rather than combined into a single
supernova likelihood. The dataset combinations used in each
analysis are specified in sets A--E below.

\subsection{Model testing}

To understand the role of the new DE model, we performed different  parameter inferences, one with fixed massive neutrinos to $0.06$ eV (a single massive and two massless neutrinos) and the other with $\sum m_\nu$ as a free parameter. We also run inferences for a flat Universe and for curvature as a free parameter (again with a single massive neutrino). The inclusion of  massive neutrinos and curvature as free parameters is motivated by  the recent debate regarding the negative mass posterior probability \cite{Craig:2024tky,Elbers:2024sha,DESI:2025ejh} and curvature as a possible solution to this issue \cite{Chen:2025mlf}.

We opted to use the same priors as in the DESI BAO DR2 cosmological analysis \cite{DESI:2025zgx} in order to compare their results for $\Lambda$CDM model with ours for the CHDE model. Our priors are displayed in Table \ref{table:priors1}, where we also vary the CHDE exponent $n$, see eq. (\ref{eq:HF2}). The CHDE strength parameter $Q$ (or  $\Omega_{\mathcal{M}}^{(0)}$)  is a derived parameter obtained from the constraint equation (\ref{eq:constraint}), similar as it happens in the $\Lambda$CDM model to determine $\Omega_\Lambda$.

For the analysis,  we consider the following set of data and models:

A) Individual data of DESI BAO DR2, Planck, Pantheon+, DESY5, and Union3. For these datasets, we perform the inference allowing spatial curvature to vary while keeping the neutrino mass fixed, since when treated as a free parameter it is better constrained with the ACT lensing information that will be included in the subsequent datasets. Results of this set are in Table \ref{tab:bao_dr2_sn_planck_individual}.

B) DESI BAO DR2 + SNe Ia for a model with free curvature and  fixed neutrino mass, as in A, but now we use two datasets for the inference.
Results of this set are in Table \ref{tab:bao_dr2_sn_combined}.

C) DESI BAO DR2 + Planck data +  SNe Ia for a model with free curvature and  fixed neutrino mass, as in A, but now we use three datasets for the inference.   Results of this set are in Table \ref{tab:bao_dr2_sn_planck_combined}.

D) DESI BAO DR2 + Planck + ACT + Union3 SNe Ia for both curvature and neutrino mass as free parameters.     Results of this set are in Table \ref{tab:bao_dr2_planck_act_dr6_non-flat}.

E) DESI BAO DR2 + Planck + ACT + Union3 SNe Ia  for a flat model and free neutrino parameter. Results of this set are in Table \ref{tab:bao_dr2_planck_act_dr6_flat}.

We clarify that Pantheon+, Union3, and DES Year~5 are treated as alternative
supernova datasets and are analyzed separately. We never
combine two or more of these compilations within the same
likelihood. For each analysis involving supernovae, we include
only one compilation, either alone or together with the BAO
and CMB datasets specified in sets A--E. In particular, the
neutrino-mass analyses in sets D and E use Union3 as the
sole supernova dataset.

For each compilation, we evaluate its respective
\texttt{Cobaya} likelihood using the CHDE luminosity-distance
prediction and the corresponding covariance matrix, including
statistical and systematic uncertainties. The overall
magnitude normalization is analytically marginalized.
For a joint analysis, the selected supernova likelihood
is multiplied by the likelihoods of the other cosmological
probes included in that dataset combination.

We emphasize that the BAO likelihoods used in this work
correspond to compressed distance measurements and therefore constrain
the background distance--redshift relation and the sound-horizon scale.
The CMB likelihoods, on the other hand, require the Boltzmann evolution
of the cosmological perturbations in the modified CHDE background. In
particular, for the analyses including massive neutrinos, where the
inferred neutrino-mass constraints may be especially sensitive to the
evolution of cosmological perturbations, we implement the covariant
energy-transfer prescription for the CHDE--dark-matter sector described
above. In the dark-matter-comoving synchronous gauge, the CHDE component carries no independent density perturbation, while the dark-matter
continuity equation is modified by the energy exchange with the CHDE
component. The perturbations of all other standard cosmological species,
including massive neutrinos, as well as the metric perturbations, are
evolved consistently within the Boltzmann calculation.

\begin{center}
\begin{table*}
\ra{1.3}
\begin{center}
\begin{tabular} { l  l }

 Parameters &  Priors  \\
\hline
\vspace{0.15cm}

$\quad n$ & $\mathcal{U}(-1, 1)$\\
\vspace{0.15cm}

$\quad \Omega_{k}$ & $\mathcal{U}(-0.3, 0.3)$\\
\vspace{0.15cm}

$\quad H_0$ $\text{[km/s/Mpc]}$ & $\mathcal{U}(20,150)$   \\
\vspace{0.15cm}

$\quad \omega_\text{b}$ & $\mathcal{G}(0.0222,0.0005^{2}) $\\
\vspace{0.15cm}

$\quad \omega_\text{cdm}$ & $\mathcal{U}(0.001,0.99) \, $\\
\vspace{0.15cm}

$\quad \ln(10^{10}A_s)$ & $\mathcal{U}(1.61,3.91)
 \, $\\

$\quad n_\text{s}$ & $\mathcal{U}(0.8,1.2)$ \\
\vspace{0.15cm}

$\quad \tau_\text{reio}$ & $\mathcal{U}(0.01,0.8)$ \\

$\quad \sum m_{\nu}$ $\text{[eV}]$ & $\mathcal{U}(0, 5)$ \\

\hline

\end{tabular}
\caption{Priors of the relevant fitting parameters. Here  \texorpdfstring{$\mathcal{U}$}{}  and \texorpdfstring{$\mathcal{G}$}{}
denote uniform and Gaussian priors, respectively. }
\label{table:priors1}
\end{center}
\end{table*}
\end{center}

The inferred cosmological parameters are summarized in the following tables and figures.
Our results report best-fitted parameters with $68\%$ confidence levels (C.L.), except
for $\Sigma m_{\nu}$ for which we report $95\%$ C.L. Table \ref{tab:bao_dr2_sn_planck_individual}  presents the results
obtained for data set A. We analyze each dataset
separately to characterize its constraining power and parameter degeneracies
within CHDE, assess consistency between probes, and identify their
contributions to the joint constraints. In particular,
uncalibrated supernova distances constrain the shape of
the distance--redshift relation but do not independently
determine $H_0$. The broad, prior-dependent $H_0$ intervals
from supernova-only analyses should be interpreted accordingly. Independent
analyses of BAO and SNe Ia allow for a relatively
broad range of parameter values within our model, indicating that the CHDE scenario remains
partially consistent with the standard $\Lambda$CDM model, which is recovered for
the specific parameter choices ($\Omega_{\mathcal{M}}^{(0)} = \Omega_\Lambda, n=0$).  However, results from  Planck data alone exclude $\Lambda$CDM. Also, the joint
analysis of BAO and SNe Ia displayed in Table \ref{tab:bao_dr2_sn_combined} disfavors
the $\Lambda$CDM model. Next, Table \ref{tab:bao_dr2_sn_planck_combined} reports the
parameter constraints derived from the combined BAO+Planck+SNe Ia data. In this case, the
$\Lambda$CDM model is disfavored with even greater statistical precision: the non-flat
model is disfavored by 4.4$\sigma$, whereas the flat model is $4.5 \sigma$ away from
$\Lambda$CDM\footnote{We quantify the separation of a specified reference
point $\boldsymbol{x}_*=(\Omega_{\mathcal M,*}^{(0)},n_*)$
from the marginalized posterior using the Gaussian approximation
$T=(\boldsymbol{\mu}-\boldsymbol{x}_*)^T
\mathbf{C}^{-1}(\boldsymbol{\mu}-\boldsymbol{x}_*)$,
where $\boldsymbol{\mu}$ and $\mathbf{C}$ are the posterior
mean and covariance in $(\Omega_{\mathcal M}^{(0)},n)$.
The equivalent two-sided Gaussian significance is
$N_\sigma=\sqrt{2}\,\mathrm{erf}^{-1}[F_{\chi^2_2}(T)]$.
This measures the separation of the specified point,
rather than a model-selection significance.}.  These results are presented
more visually in Figure \ref{fig:figUre1}, which shows contour plots for the best-fitted
parameters of the model ($\Omega_{\mathcal{M}}^{(0)}$, $n$) for the analysis combining
the different sets. The joint analysis clearly excludes the $\Lambda$CDM model (dashed
lines) as the best fit. Note that the apparent position of the joint BAO+CMB contour
should be interpreted with some care, since it is obtained from the full multidimensional
joint likelihood and then projected onto the $(\Omega_{\mathcal M}^{(0)},n)$ plane.
Therefore, it does not necessarily coincide with the visual overlap of the
independently marginalized BAO and CMB contours.

\begin{figure*}
\centering
\includegraphics[width=0.45\textwidth]{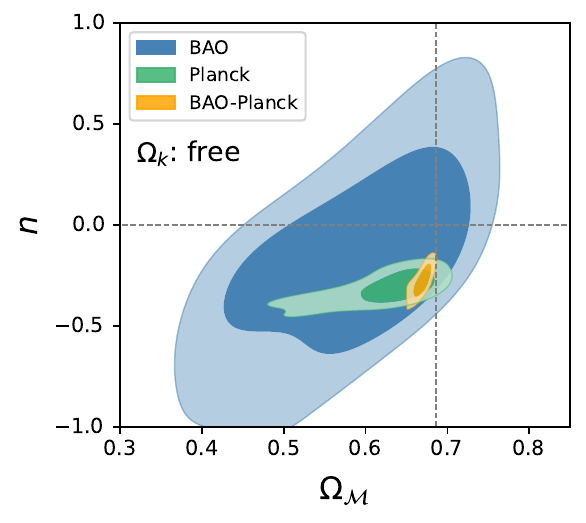}
\includegraphics[width=0.45\textwidth]{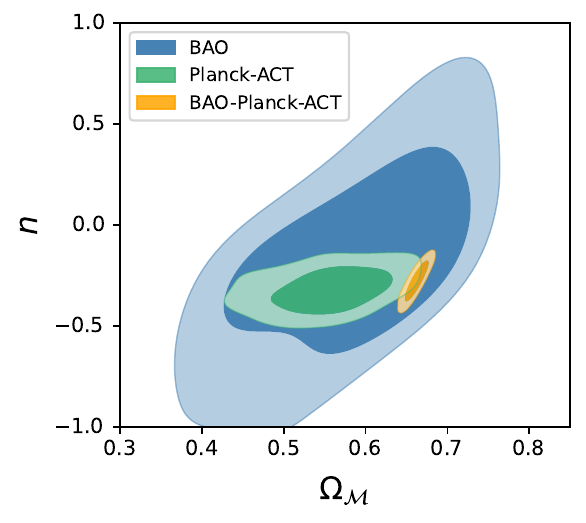}

\vspace{0.3cm}

\includegraphics[width=0.45\textwidth]{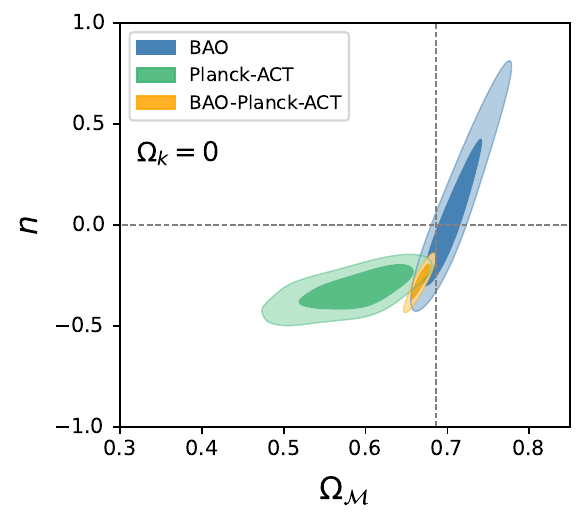}
\caption{Contour plots of the CHDE model parameters: the dark-energy density parameter $\Omega_{\mathcal{M}}^{(0)}$ and the power-law exponent $n$, cf. Eq.~(\ref{eq:HF2}). The left panel shows BAO, Planck, and BAO+Planck constraints with curvature as a free parameter. The upper-right panel shows BAO, Planck+ACT, and BAO+Planck+ACT constraints, also allowing curvature to vary. The lower panel shows the corresponding BAO, Planck+ACT, and BAO+Planck+ACT constraints for the flat model. Our results exclude the $\Lambda$CDM limit, $\Omega_{\mathcal{M}}^{(0)}=\Omega_\Lambda$ and $n=0$, as the best-fit solution when CMB data are included.}
\label{fig:figUre1}
\end{figure*}

As it happens in the $\Lambda$CDM model, $\Omega_m$ is smaller for BAO than for Planck, but the discrepancy is larger in our CHDE model. Figure \ref{fig:figUre_Omega_m}  shows the matter density parameter vs the Hubble constant (left panel) and 1-dimensional (1D) marginalized posteriors when using different datasets (right panel).  Also in this figure one observes a shift in the value of the Hubble constant towards larger values than in $\Lambda$CDM (vertical dashed line). The reported value in Table \ref{tab:bao_dr2_sn_planck_combined} for BAO-Planck-Pantheon+ data is  $H_0 = 70.75^{+0.81}_{- 1.2} $ km$s^{-1}$  ${\rm Mpc}^{-1}$ when $\Omega_k$ is a free parameter, whereas in $\Lambda$CDM is $63.3 \pm 2.1$ km$s^{-1}$  ${\rm Mpc}^{-1}$ \cite{DESI:2025zgx}. For a flat cosmology, we report in Table \ref{tab:bao_dr2_planck_act_dr6_flat}  a value of  $67.40 \pm  0.36$ km$s^{-1}$  ${\rm Mpc}^{-1}$  and $\Lambda$CDM for Planck data is   $67.36 \pm 0.54$ km$s^{-1}$   ${\rm Mpc}^{-1}$ \cite{Planck:2018vyg}, that are compatible values to 1$\sigma$. These results are  shown in Figure \ref{fig:figUre_H0}. Thus, the results for $H_0$ do not solve the Hubble tension using early universe physics (CMB and BAO).

\begin{figure*}
	\centering
	\includegraphics[width=0.45\textwidth]{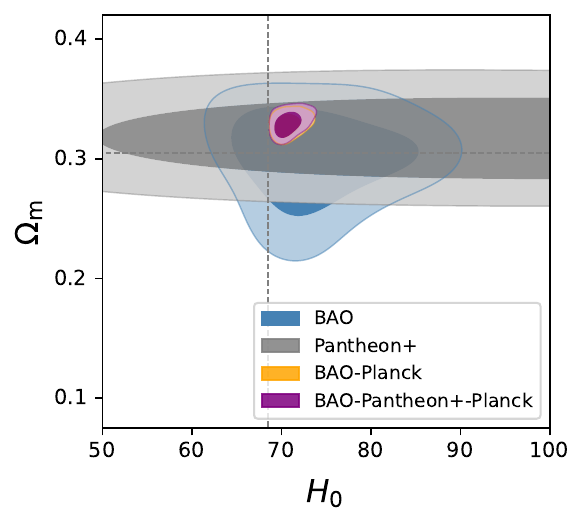}
    \includegraphics[width=0.45\textwidth]{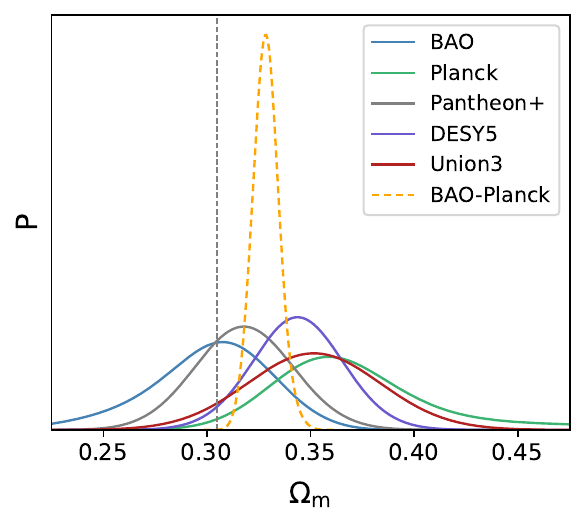}
\caption{Contour plots of matter density parameter vs the Hubble constant for the CHDE model (left) and the  marginalized 1D posteriors probabilities (P) for different datasets (right). As a reference, the black dashed line points to the best fit values for the \texorpdfstring{$\Lambda$}{}CDM model using Planck data.} \label{fig:figUre_Omega_m}
\end{figure*}

\begin{figure*}
	\centering
	\includegraphics[width=0.60\textwidth]{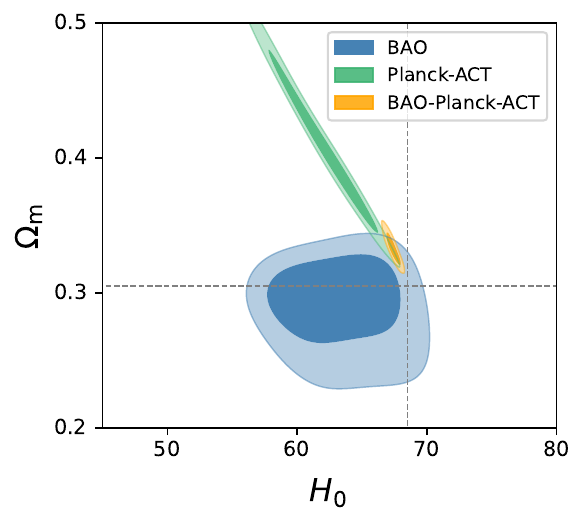}
\caption{Contour plots of matter density parameter vs the Hubble constant for the flat CHDE model. As a reference, the black dashed line points to the best fit values for the \texorpdfstring{$\Lambda$}{}CDM model using Planck data.} \label{fig:figUre_H0}
\end{figure*}

With runs A-D we explore the role of curvature in our CHDE model. The curvature behaves similar to the $\Lambda$CDM model's:  Planck fitting results in a small negative $\Omega_k$ \cite{Rosenberg:2022sdy}, whereas DESI BAO fitting favors a positive value \cite{DESI:2025zgx}, although in our case $\Omega_k$ is larger and has larger uncertainties, since we include an additional parameter.  These results are shown in Table \ref{tab:bao_dr2_sn_planck_individual},  and  Figure \ref{fig:figUre_Omega_k} shows the contour plots of the curvature vs the matter parameter. Further fittings that include BAO data in Tables \ref{tab:bao_dr2_sn_combined}-\ref{tab:bao_dr2_planck_act_dr6_non-flat} show a preference for a positive $\Omega_k$.  The low $H_0$ values inferred from Planck+ACT illustrate
the model dependence of the CMB-based determination of
the present expansion rate. The modified CHDE evolution,
together with free neutrino mass and spatial curvature
where applicable, permits parameter combinations with
low $H_0$ and higher matter density. Since low $H_0$
also occurs in the flat analysis, curvature is not the
sole explanation. Adding BAO constrains the distance--redshift
relation at several epochs, restricts these degeneracies,
and shifts the posterior toward higher $H_0$ with smaller
uncertainties. We do not attribute this shift to a specific
observational systematic.

In recent determinations of the summed neutrino mass ($\sum m_\nu$), the DESI collaboration reports very small values that may eventually come into tension with terrestrial
oscillation experiments, which establish a lower bound on this quantity of $\sum m_\nu > 0.059$ eV for normal mass
hierarchy (NH) and $\sum m_\nu > 0.1$ eV for inverted hierarchy (IH) \cite{Esteban:2024eli}. The result found in
the DESI DR2 analysis is, for a flat cosmology, $\sum m_\nu < 0.0642$ eV ($95 \%$ C.L.) \cite{DESI:2025zgx}, and SPT-3G D1+DESI and CMB-SPA+DESI combinations appear to rule out
the normal and inverted hierarchies at $97.9\%$ and $99.9 \%$ confidence, respectively \cite{SPT-3G:2025bzu}. Furthermore, the maximum posterior probability lies at nonphysical, negative values \cite{DESI:2025ejh}. These conflicting results open the possibility for alternative scenarios. In this sense, reference \cite{Chen:2025mlf} argues that inclusion of the curvature as free parameter relaxes the constraint to  $\sum m_\nu < 0.1$ eV ($95 \%$ C.L.) and makes the posterior probability more positive, suggesting that curvature may alleviate the problem. However, their maximum probability is around zero, suggesting that a full resolution  would require performing an analysis with effective negative neutrino masses \cite{DESI:2025ejh,Pulido-Hernandez:2026hcs}.
In our CHDE model we computed the neutrino mass, using sets D (free curvature) and E (flat model), obtaining the results in Tables  \ref{tab:bao_dr2_planck_act_dr6_non-flat} and \ref{tab:bao_dr2_planck_act_dr6_flat}: we obtain $\sum m_\nu = 0.34^{+0.34}_{-0.30}$ eV ($95 \%$ C.L.) with a negative $\Omega_k$ when using Planck+ACT data and $\sum m_\nu = 0.037^{+0.070}_{-0.052}$ eV ($95 \%$ C.L.) with BAO+Planck+ACT data resulting in a positive $\Omega_k$ as in ref. \cite{Chen:2025mlf}; in addition, we performed the inference including the Union3 SNe Ia data set, since supernovae are known to shift the neutrino mass toward negative values. We find that, in the CHDE model, the neutrino mass remains positive, with a significance similar to that obtained with the previous data sets.
For a flat cosmology, we obtain $\sum m_\nu = 0.39^{+0.46}_{-0.39}$ eV ($95 \%$ C.L.) when  using Planck+ACT data and $\sum m_\nu = 0.021^{+0.040}_{-0.029}$ eV ($95 \%$ C.L.) with BAO+Planck+ACT data; we further included the Union3 SNe Ia data set and found similar results, with the neutrino mass remaining positive. All these results are compatible with the terrestrial oscillation experiment constraints within $95\%$ C.L.  Furthermore, Figure \ref{fig:figUre_mnu_posteriors} depicts the marginalized 1D posterior probabilities for non-flat and flat models, clearly showing that, in both cases, the maximum lies at positive values of $\sum m_\nu$ in the CHDE model. The flat $\Lambda$CDM model provides a clearer indication against the inverted hierarchy (IH). On the other hand, as discussed above, DESI data favor a lower value of
$\Omega_m$ than Planck data. This mismatch reduces the overlap
between the parameter regions preferred by the two datasets when
they are combined, particularly in the $\sum m_\nu$ direction.
Figure~\ref{fig:figUre_neutrino_Omega_m} shows these results for
both flat and non-flat models. The concordance region, where the
probability contours overlap, is relatively small, which drives
the combined constraint toward a small value of $\sum m_\nu$. We caution that the perturbation model used here is a particular one, and that different perturbation settings can change the neutrino mass results, as shown in ref. \cite{Yang:2026kxp}.

\begin{figure*}
	\centering
	\includegraphics[width=0.60\textwidth]{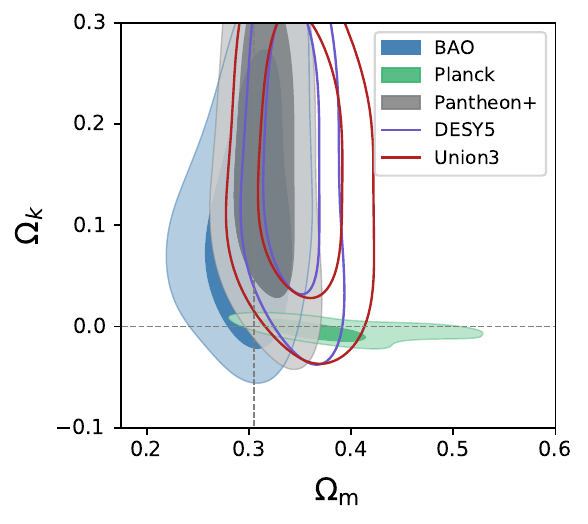}
\caption{Contour plots of the curvature and matter density parameters for the CHDE model.} \label{fig:figUre_Omega_k}
\end{figure*}

\begin{figure*}
	\centering
	\includegraphics[width=0.45\textwidth]{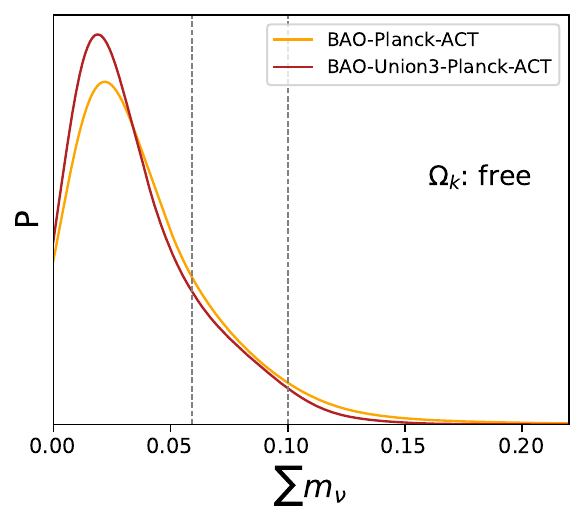}
    \includegraphics[width=0.45\textwidth]{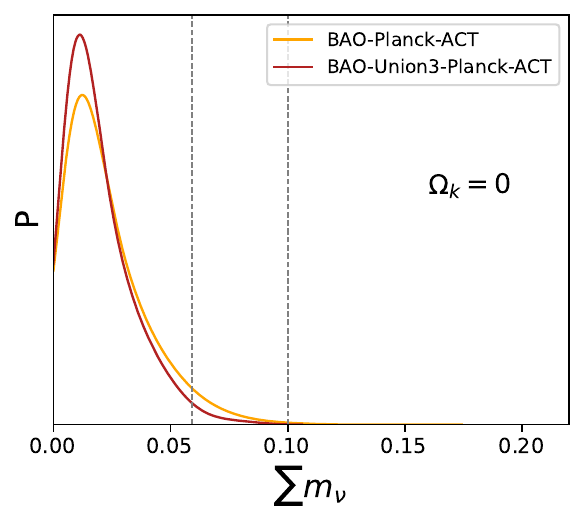}
\caption{Posterior probabilities of the sum of the neutrino masses for non-flat (left panel) and flat (right panel) models for the CHDE model. Both plots show a maximum probability in the positive neutrino masses. For reference vertical dashed lines indicate the NH and IH experimental limits from terrestrial experiments, at \texorpdfstring{$0.059$}{} eV and \texorpdfstring{$0.1$}{}  eV, respectively.} \label{fig:figUre_mnu_posteriors}
\end{figure*}

\begin{figure*}
	\centering
	\includegraphics[width=0.45\textwidth]{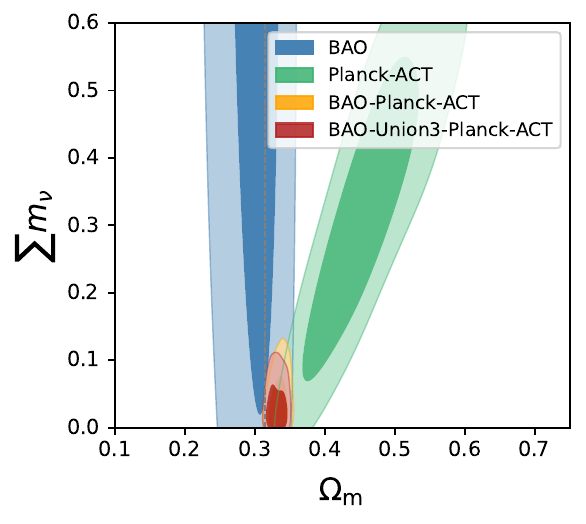}
    \includegraphics[width=0.45\textwidth]{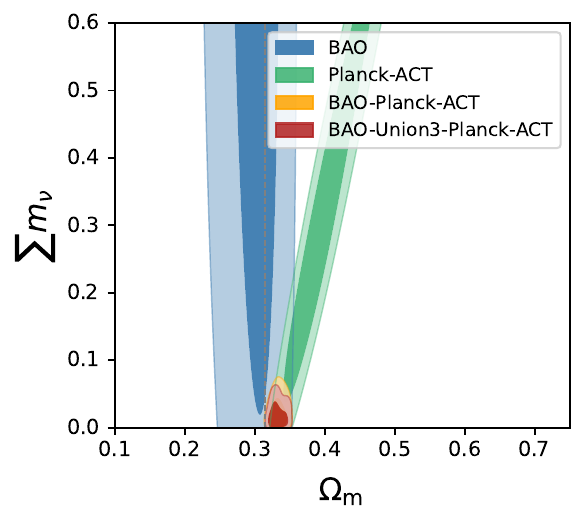}
\caption{Contour plots of the neutrino mass vs the matter density parameter. Left panel is for \texorpdfstring{$\Omega_k$}{} as a free parameter and right panel for a flat model. The region of concordance in the joint analysis is quite narrow, driving the inferred neutrino mass towards very small values.} \label{fig:figUre_neutrino_Omega_m}
\end{figure*}

\begin{table}

\centering
\scriptsize
\setlength{\tabcolsep}{4pt}
\renewcommand{\arraystretch}{1.15}

\resizebox{\textwidth}{!}{
\begin{tabular}{lccccc}
\hline\hline
 & \textbf{BAO} & \textbf{Planck} & \textbf{Pantheon+} & \textbf{DESY5} & \textbf{Union3} \\
\cline{2-6}
\textbf{Parameter} & 68\% limits & 68\% limits & 68\% limits & 68\% limits & 68\% limits \\
\hline
$n$
& $-0.15^{+0.31}_{-0.37}$
& $-0.312\pm 0.056$
& $-0.28^{+0.30}_{-0.49}$
& $-0.41^{+0.22}_{-0.50}$
& $-0.40^{+0.18}_{-0.55}$ \\

$\Omega_k$
& $0.107^{+0.069}_{-0.10}$
& $-0.0041^{+0.0060}_{-0.0068}$
& $0.149^{+0.11}_{-0.077}$
& $0.151^{+0.10}_{-0.075}$
& $0.148^{+0.098}_{-0.078}$ \\

$\Omega_\mathrm{b}h^2$
& $0.02222\pm 0.00050$
& $0.02226^{+0.00041}_{-0.00017}$
& $0.02222\pm 0.00050$
& $0.02222\pm 0.00050$
& $0.02223\pm 0.00050$ \\

$\Omega_\mathrm{m}$
& $0.302^{+0.033}_{-0.025}$
& $0.374^{+0.023}_{-0.048}$
& $0.318\pm 0.024$
& $0.344\pm 0.022$
& $0.352\pm 0.032$ \\

$H_0\,[\mathrm{km\,s^{-1}\,Mpc^{-1}}]$
& $73.7^{+4.6}_{-7.5}$
& $66.7\pm 3.5$
& $108^{+40}_{-40}$
& $109^{+40}_{-40}$
& $109^{+40}_{-40}$ \\

$\Omega_{\mathcal{M}}$
& $0.592^{+0.12}_{-0.073}$
& $0.630^{+0.045}_{-0.019}$
& $0.533\pm 0.084$
& $0.505\pm 0.082$
& $0.500^{+0.085}_{-0.094}$ \\
\hline
\end{tabular}
}
\caption{Parameter inference for the relevant cosmological parameters using independent BAO DESI DR2, Planck, and SNe Ia data. Results correspond to set A, as explained in the main text.}
\label{tab:bao_dr2_sn_planck_individual}

\end{table}

\begin{table*}
\small
    \centering

\begin{tabular} { l  c c c}
\noalign{\vskip 3pt}\hline\noalign{\vskip 1.5pt}\hline\noalign{\vskip 5pt}
 \multicolumn{1}{c}{\bf } &  \multicolumn{1}{c}{\bf BAO-Pantheon+} &  \multicolumn{1}{c}{\bf BAO-DESY5} &  \multicolumn{1}{c}{\bf BAO-Union3}\\
\noalign{\vskip 3pt}\cline{2-4}\noalign{\vskip 3pt}

 Parameter &  68\% limits &  68\% limits &  68\% limits\\
\hline
{$n              $} & $-0.30\pm 0.24             $ & $-0.52\pm 0.24             $ & $-0.44\pm 0.28             $\\

{$\Omega_k       $} & $0.129^{+0.080}_{-0.10}    $ & $0.156\pm 0.078            $ & $0.146\pm 0.081            $\\

{$\Omega_\mathrm{b} h^2$} & $0.02222\pm 0.00050        $ & $0.02223\pm 0.00050        $ & $0.02223\pm 0.00050        $\\

$\Omega_\mathrm{m}         $ & $0.314\pm 0.019            $ & $0.332\pm 0.017            $ & $0.326\pm 0.021            $\\

$H_\mathrm{0} $\text{[km/s/Mpc]}$             $ & $75.0^{+6.2}_{-7.9}        $ & $74.3^{+6.9}_{-8.0}        $ & $74.4^{+6.7}_{-8.0}        $\\

$\Omega_{\mathcal{M}}      $ & $0.556^{+0.10}_{-0.073}    $ & $0.511\pm 0.081            $ & $0.528\pm 0.086            $\\
\hline
\end{tabular}

    \caption{Results for the joint analysis of BAO DR2 + SNe Ia data. Results from set B.
    \label{tab:bao_dr2_sn_combined}
    }
\end{table*}

\begin{table*}
\small
    \centering

\begin{tabular} { l  c c c}
\noalign{\vskip 3pt}\hline\noalign{\vskip 1.5pt}\hline\noalign{\vskip 5pt}
 \multicolumn{1}{c}{\bf } &  \multicolumn{1}{c}{\bf BAO-Pantheon+-Planck} &  \multicolumn{1}{c}{\bf BAO-DESY5-Planck} &  \multicolumn{1}{c}{\bf BAO-Union3-Planck}\\
\noalign{\vskip 3pt}\cline{2-4}\noalign{\vskip 3pt}

 Parameter &  68\% limits &  68\% limits &  68\% limits\\
\hline
{$n              $} & $-0.285\pm 0.061           $ & $-0.284\pm 0.052           $ & $-0.280\pm 0.054           $\\

{$\Omega_k       $} & $0.0027^{+0.0016}_{-0.0033}$ & $0.0021^{+0.0015}_{-0.0024}$ & $0.0023^{+0.0016}_{-0.0024}$\\

{$\Omega_\mathrm{b} h^2$} & $0.02250^{+0.00018}_{-0.00015}$ & $0.02249\pm 0.00016        $ & $0.02248\pm 0.00016        $\\

$\Omega_\mathrm{m}         $ & $0.3295^{+0.0066}_{-0.0076}$ & $0.3304\pm 0.0063          $ & $0.3296^{+0.0062}_{-0.0070}$\\

$H_\mathrm{0}  $\text{[km/s/Mpc]}$            $ & $70.75^{+0.81}_{-1.2}      $ & $70.43^{+0.74}_{-0.85}     $ & $70.43^{+0.70}_{-0.87}     $\\

$\Omega_{\mathcal{M}}      $ & $0.6678^{+0.0096}_{-0.0068}$ & $0.6674^{+0.0074}_{-0.0063}$ & $0.6680^{+0.0081}_{-0.0064}$\\
\hline
\end{tabular}

    \caption{Results for the joint analysis employing BAO DR2+Planck+SNe Ia data. Results from set C.
    \label{tab:bao_dr2_sn_planck_combined}
    }
\end{table*}

\begin{table*}
\small
    \centering

\begin{tabular} { l  c c c}
\noalign{\vskip 3pt}\hline\noalign{\vskip 1.5pt}\hline\noalign{\vskip 5pt}
 \multicolumn{1}{c}{\bf } &  \multicolumn{1}{c}{\bf Planck-ACT} &  \multicolumn{1}{c}{\bf BAO-Planck-ACT} &  \multicolumn{1}{c}{\bf BAO-Union3-Planck-ACT}\\
\noalign{\vskip 3pt}\cline{2-4}\noalign{\vskip 3pt}

 Parameter &  68\% limits &  68\% limits &  68\% limits\\
\hline
{$n              $} & $-0.321\pm 0.080           $ & $-0.279\pm 0.064           $ & $-0.239\pm 0.072           $\\

{$\Omega_k       $} & $-0.0176\pm 0.0079         $ & $0.0038^{+0.0017}_{-0.0022}$ & $0.0032^{+0.0015}_{-0.0018}$\\

{$\Omega_\mathrm{b} h^2$} & $0.02215\pm 0.00016        $ & $0.02220\pm 0.00013        $ & $0.02219\pm 0.00013        $\\

$\Omega_\mathrm{m}         $ & $0.462^{+0.050}_{-0.059}   $ & $0.3344\pm 0.0086          $ & $0.3310\pm 0.0091          $\\

$H_\mathrm{0}              $ & $57.8\pm 3.4               $ & $67.68\pm 0.39             $ & $67.70\pm 0.40             $\\

$\Omega_{\mathcal{M}}      $ & $0.556^{+0.052}_{-0.044}   $ & $0.6618\pm 0.0096          $ & $0.666^{+0.011}_{-0.0098}  $\\

$\sum m_\nu       $\text{[eV}]$ \, (95\% \, {\rm C.L.})         $ & $0.34^{+0.34}_{-0.30}      $ & $0.037^{+0.070}_{-0.052}   $ & $0.033^{+0.062}_{-0.044}   $\\
\hline
\end{tabular}

\caption{Results for the joint analysis employing Planck+ACT DR6 and BAO DR2+Planck+ACT DR6 data for a non-flat model. Neutrino masses are given in eV and  reported with 95\% C.L., as standard. Results from set D.
    \label{tab:bao_dr2_planck_act_dr6_non-flat}
    }
\end{table*}

\begin{table*}
\small
    \centering

\begin{tabular} { l  c c c}
\noalign{\vskip 3pt}\hline\noalign{\vskip 1.5pt}\hline\noalign{\vskip 5pt}
 \multicolumn{1}{c}{\bf } &  \multicolumn{1}{c}{\bf Planck-ACT} &  \multicolumn{1}{c}{\bf BAO-Planck-ACT} &  \multicolumn{1}{c}{\bf BAO-Union3-Planck-ACT}\\
\noalign{\vskip 3pt}\cline{2-4}\noalign{\vskip 3pt}

 Parameter &  68\% limits &  68\% limits &  68\% limits\\
\hline
{$n              $} & $-0.322\pm 0.074           $ & $-0.279^{+0.065}_{-0.052}  $ & $-0.262^{+0.089}_{-0.11}   $\\

{$\Omega_\mathrm{b} h^2$} & $0.02196^{+0.00021}_{-0.00017}$ & $0.02233\pm 0.00012        $ & $0.02231\pm 0.00012        $\\

$\Omega_\mathrm{m}         $ & $0.412^{+0.036}_{-0.054}   $ & $0.3327^{+0.0073}_{-0.0086}$ & $0.3320^{+0.0062}_{-0.011} $\\

$H_\mathrm{0}              $ & $61.9^{+3.3}_{-2.4}        $ & $67.40\pm 0.36             $ & $67.37^{+0.38}_{-0.33}     $\\

$\Omega_{\mathcal{M}}      $ & $0.588^{+0.054}_{-0.036}   $ & $0.6672^{+0.0086}_{-0.0073}$ & $0.668^{+0.011}_{-0.0062}  $\\

$\sum m_\nu  $\text{[eV}]$ \, (95\% \, {\rm C.L.})               $ & $0.39^{+0.46}_{-0.39}      $ & $0.021^{+0.040}_{-0.029}   $ & $0.018^{+0.033}_{-0.024}$\\
\hline
\end{tabular}

    \caption{Results for the joint analysis employing Planck+ACT DR6 and BAO DR2+Planck+ACT DR6 data for a flat model. Neutrino masses are given in eV and  reported with 95\% C.L., as standard. Results from set E.
    \label{tab:bao_dr2_planck_act_dr6_flat}
    }
\end{table*}

Figures~\ref{fig:figUre_full_contours1}--\ref{fig:figUre_full_contours3}
present the marginalized posterior distributions and joint
constraints on the cosmological parameters for sets C--E.
Figure~\ref{fig:figUre_full_contours1} shows the results for
set C, in which spatial curvature varies and the summed
neutrino mass is fixed to $0.06\,\mathrm{eV}$. The different
BAO+Planck+SNe~Ia combinations yield mutually consistent
constraints, with a preference for a negative CHDE exponent
$n$ and a small positive curvature parameter $\Omega_k$.
Each combination includes only one supernova compilation.

Figure~\ref{fig:figUre_full_contours2} presents the results
for set D, allowing both spatial curvature and the summed
neutrino mass to vary.
The comparison between Planck+ACT and the combinations
including BAO illustrates the substantial tightening and
shifting of the parameter constraints when BAO information
is included, particularly for $\Omega_m$, $H_0$, $\Omega_k$,
and $\sum m_\nu$.
Figure~\ref{fig:figUre_full_contours3} shows the corresponding
results for set E, which assumes spatial flatness while
allowing the summed neutrino mass to vary.
In both sets, adding BAO shifts the neutrino-mass constraints
toward smaller values, while the further inclusion of Union3
produces comparatively modest changes.
The full distributions complement the selected parameter
projections discussed above by displaying the correlations
and the dependence of the inferred constraints on the
dataset combination and the assumed spatial geometry.

\begin{figure*}
	\centering
	\includegraphics[width=1 \textwidth]{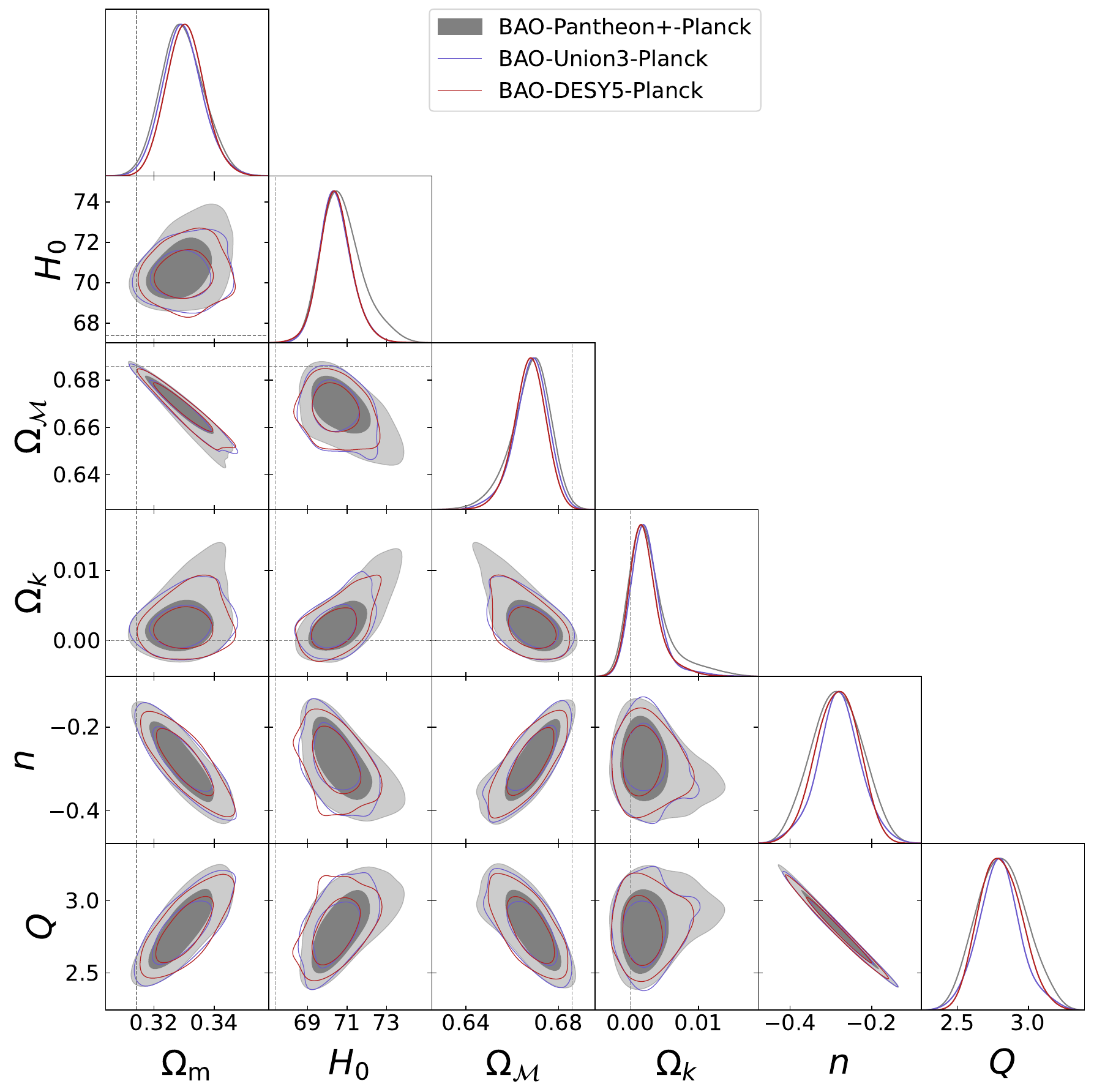}
\caption{Contour plots of the relevant parameters for the CHDE model using set C with a free curvature model.}  \label{fig:figUre_full_contours1}
\end{figure*}

\begin{figure*}
	\centering
	\includegraphics[width=1 \textwidth]{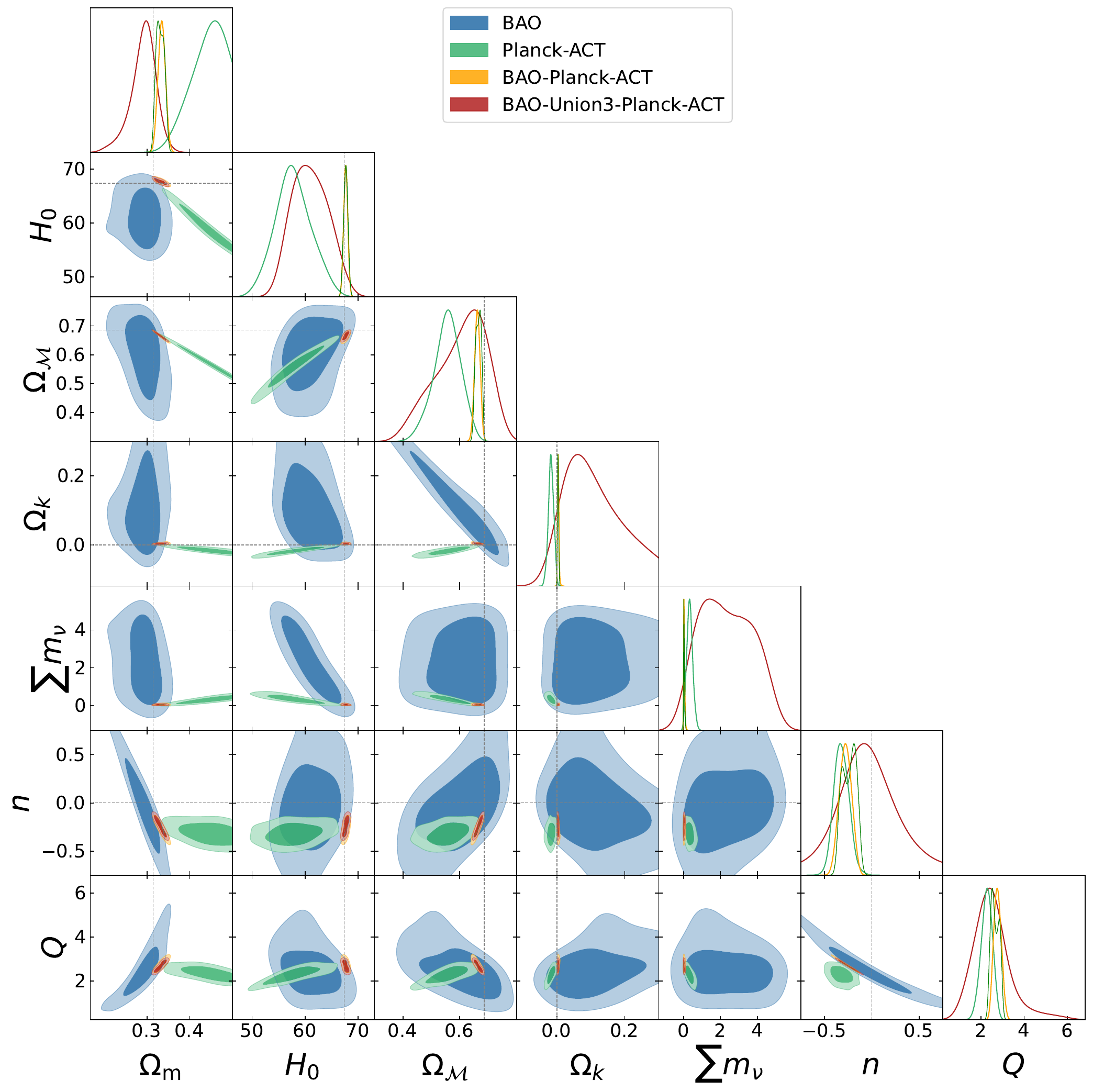}
\caption{Contour plots of the relevant parameters for the CHDE model using set D with a free curvature model and free neutrino mass.}  \label{fig:figUre_full_contours2}
\end{figure*}

\begin{figure*}
	\centering
	\includegraphics[width=1 \textwidth]{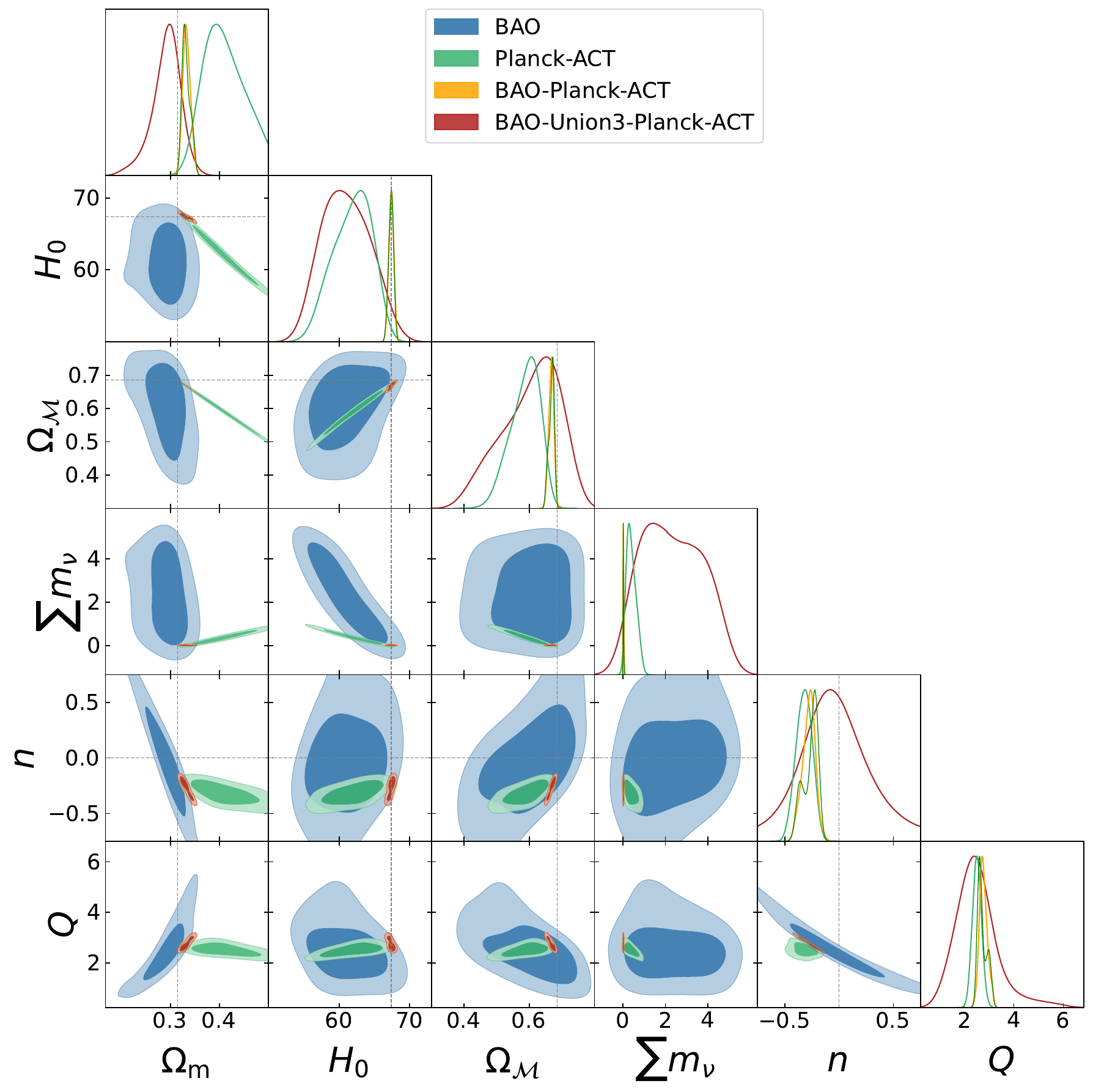}
\caption{Contour plots of the relevant parameters for the CHDE model using set E for a flat model and free neutrino mass.}  \label{fig:figUre_full_contours3}
\end{figure*}

\subsection{Goodness of fit}

Finally, to assess model performance beyond parameter estimation
and compare their goodness of fit and statistical support, we
computed three statistical metrics using the maximum a posteriori (MAP)
solutions obtained with Cobaya's \texttt{BOBYQA} optimizer. Specifically,
we report $\Delta\chi^2_{\mathrm{MAP}}$, together with MAP-based
estimates of the Akaike Information Criterion (AIC) and the
Bayesian Information Criterion (BIC). We follow the expressions
given in \cite{Liddle:2007fy}, but evaluate the likelihood at
the MAP rather than at its maximum.
Table \ref{goodness-of-fit} presents the results for the CHDE
model with neutrinos, with and without spatial curvature,
relative to the corresponding $\Lambda$CDM model. For each criterion $X_i$, we define
$\Delta X_i \equiv X_i^{\mathrm{CHDE}}
- X_i^{\Lambda\mathrm{CDM}}$, so that negative values favor
CHDE. The CHDE model with curvature provides a slightly better
fit according to $\Delta\chi^2_{\mathrm{MAP}}$ and is mildly
preferred by the $\Delta$AIC, whereas the
$\Delta$BIC favors $\Lambda$CDM. The flat CHDE model provides
a comparable fit to $\Lambda$CDM according to
$\Delta\chi^2_{\mathrm{MAP}}$, while both the
$\Delta$AIC and $\Delta$BIC favor $\Lambda$CDM.

\begin{table}[htbp]
\centering
\caption{Goodness-of-fit comparison between CHDE and
$\Lambda$CDM. Differences are CHDE minus $\Lambda$CDM.}

\label{goodness-of-fit}
\begin{tabular}{lll}
\hline
Criterion & Difference & Assessment \\
\hline
\multicolumn{3}{c}{With curvature} \\
\hline
$\Delta\chi^2_{\mathrm{MAP}}$ & $-2.161$
& CHDE slightly better fit \\
$\Delta\mathrm{AIC}$ & $-0.161$
& CHDE slightly preferred \\
$\Delta\mathrm{BIC}$ & $+7.05$
& $\Lambda$CDM preferred \\
\hline
\multicolumn{3}{c}{Without curvature} \\
\hline
$\Delta\chi^2_{\mathrm{MAP}}$ & $+0.056$
& Essentially identical fits \\
$\Delta\mathrm{AIC}$ & $+2.056$
& $\Lambda$CDM preferred \\
$\Delta\mathrm{BIC}$ & $9.262$
& $\Lambda$CDM preferred \\
\hline
\end{tabular}
\end{table}

\subsection{Physical interpretation}
The preference for negative $n$ has a direct physical
interpretation. Since $\rho_{\mathcal M}\propto\eta^n$,
its conformal-time derivative is
$\rho'_{\mathcal M}=n\rho_{\mathcal M}/\eta$.
Thus, for positive dark-energy density and $n<0$,
the CHDE density decreases as the Universe evolves.
The dark-matter continuity equation,
\begin{equation}
\rho'_{\mathrm{dm}}+3\mathcal H\rho_{\mathrm{dm}}
=-\rho'_{\mathcal M}
=-\frac{n\rho_{\mathcal M}}{\eta},
\end{equation}
then contains a positive source term, describing energy
transfer from CHDE to dark matter. Equivalently,
\begin{equation}
\left(a^3\rho_{\mathrm{dm}}\right)'
=-a^3\rho'_{\mathcal M}>0.
\end{equation}
The dark-matter energy in a fixed comoving volume
therefore increases. This does not necessarily imply
that its physical density increases: expansion can
still dilute dark matter, but less rapidly than the
$a^{-3}$ scaling of a separately conserved pressureless
component.
The inferred negative exponent consequently identifies
a decaying vacuum-like dark-energy component that sources
dark matter while retaining $w_{\mathcal M}=-1$.
Its fractional contribution can nevertheless grow and
dominate at late times because the other components
dilute more rapidly. In the limit $n=0$, the transfer
vanishes and the standard $\Lambda$CDM evolution is recovered. \\
The interaction also affects dark-matter perturbations.
In the dark-matter-comoving synchronous gauge, their
fractional density contrast obeys
\begin{equation}
\delta'_{\mathrm{dm}}
=-\frac{h'}{2}
+\frac{n\rho_{\mathcal M}}{\eta\rho_{\mathrm{dm}}}
\delta_{\mathrm{dm}}.
\end{equation}
For $n<0$, the interaction term opposes the growth of
overdensities. In the adopted prescription,
$\delta\rho_{\mathcal M}=0$ in this gauge, so the energy
transfer increases the mean dark-matter density without
a corresponding perturbed source, thereby diluting
the fractional contrast. This contribution does not
necessarily make $\delta_{\mathrm{dm}}$ decrease with
time, since gravitational growth can still dominate. This dilution suggests a tendency toward suppressed matter clustering and a lower $\sigma_8$, potentially also reducing $ S_8 \equiv \sigma_8\sqrt{\Omega_m/0.3}$, although the net effect depends on the coupled background and perturbation evolution and the inferred cosmological parameters. A dedicated analysis of these quantities and growth-rate observables is left for future work.

\section{Conclusions}
\label{conclu}

The striking recent results put forward by the DESI collaboration
\cite{DESI:2024mwx,DESI:2025zgx} that suggest an evolving DE component, instead of a
constant, motivate the exploration of alternative models for the background expansion. Furthermore, notable discrepancies were identified in the inferred values of the
matter density parameter, for which DESI yields a lower value, while Planck favors a higher
one. Although the neutrino mass determined by DESI appears to be within the range allowed by
oscillation experiments, the resulting posterior lies on the negative-mass side.

As an alternative to $\Lambda$CDM, we propose a model, the conformal holographic dark energy model, dubbed CHDE, in which dark energy  depends on the conformal time through an exponent $n$, which can be constrained by observations; this represents an extra parameter with respect to the $\Lambda$CDM model. The model differs from the standard holographic formulation but resembles the vacuum dark energy model \cite{Wands:2012vg}; its equation of state is $w_{DE} =-1$, but it includes an interaction between dark matter and dark energy through the horizon term. We have tested the model using different datasets of  BAO and SNe Ia distances and CMB spectra, which were also employed by the DESI collaboration in their recent results.  We divided our parameter inference runs into different sets (A-E) to address different questions. First of all, our model theoretically includes the $\Lambda$CDM model for the case $n=0$. Our results, however, demonstrate that $\Lambda$CDM is not favored as a best fit when using CMB data alone or in joint analysis with BAO and SNe Ia data. Tables  \ref{tab:bao_dr2_sn_planck_combined}-\ref{tab:bao_dr2_planck_act_dr6_flat} show consistent results  for flat and non-flat models separately: Using Planck+ACT data alone results in $n \sim -0.32 \pm 0.1$ and using BAO+Planck+SNe Ia and BAO+Planck+ACT predicts approximately $n \sim -0.28 \pm 0.1$. In general all our results are consistent with a negative value of the exponent that implies, from eq. (\ref{eq:conserva}), a positive source of matter production, implying that our CHDE model is in fact a model with  dark energy density that begins to dominate, but decays with time following the power law, eq. (\ref{eq:HF2}). The energy exchange also modifies dark-matter perturbations. For the preferred negative values of $n$, it introduces a damping term in the evolution equation for the fractional
density contrast, opposing gravitational growth.

As in $\Lambda$CDM, DESI favors a lower $\Omega_m$ than
Planck within the CHDE model. The limited overlap between
their preferred parameter regions drives the combined
constraints toward small neutrino masses, as illustrated
in Figure~\ref{fig:figUre_neutrino_Omega_m}.
In both flat and non-flat analyses, our neutrino-mass
constraints remain compatible with terrestrial oscillation
bounds, and the posterior distributions peak at positive
values of $\sum m_\nu$, rather than exhibiting the
negative-mass preference reported in some $\Lambda$CDM
analyses.These results may depend on the adopted dark-energy
perturbation prescription, since alternative treatments
can shift the inferred neutrino masses\cite{Yang:2026kxp}.

Finally, while CHDE provides neutrino-mass constraints compatible with terrestrial oscillation bounds, its
statistical comparison with $\Lambda$CDM yields mixed
results. With spatial curvature free, CHDE provides a
slightly better fit and is mildly preferred by the AIC, whereas the BIC favors $\Lambda$CDM. For flat cosmologies, the fits are comparable and both information criteria favor $\Lambda$CDM. Thus, these comparisons provide no consistent statistical preference
for CHDE over $\Lambda$CDM.

\section*{Author contributions}
All authors contributed to all sections of this work in a similar manner.

\section*{Funding}
The authors acknowledge support by SECIHTI project CBF2023-2024-589.

\section*{Institutional review}
Not applicable.

\section*{Informed consent}
Not applicable.


\section*{Acknowledgments}

The authors thank Hernán E. Noriega for helpful comments on the neutrino section. MARM acknowledges that the results and analysis of MCMC chains in this work used the
DiRAC@Durham facility managed by the Institute for Computational Cosmology on behalf
of the STFC DiRAC HPC Facility (www.dirac.ac.uk). The equipment was funded by BEIS
capital funding via STFC capital grants ST/K00042X/1, ST/P002293/1, ST/R002371/1 and
ST/S002502/1, Durham University and STFC operations grant ST/R000832/1. DiRAC is
part of the National e-Infrastructure in the U.K.

\section*{Conflicts of interest}
The authors declare no conflicts of interest.

\section*{Abbreviations}
{
The following abbreviations are used in this manuscript:\\

\noindent
\begin{tabular}{@{}ll}
ACT & Atacama Cosmology Telescope \\
BAO & Baryon Acoustic Oscillations \\
CHDE & Conformal Holographic Dark Energy \\
C.L. & Confidence Levels \\
\texttt{class} & Cosmic Linear Anisotropy Solving System \\
CMB & Cosmic Microwave Background \\
DE & Dark Energy\\
DES & Dark Energy Survey\\
DESI & Dark Energy Spectroscopic Instrument \\
$\Lambda$CDM & $\Lambda$ Cold Dark Matter \\
MCMC & Markov chain Monte Carlo \\
SNe Ia & Supernovae Type Ia
\end{tabular}
}

\bibliographystyle{unsrtnat}

\bibliography{CHDE}

\appendix
\numberwithin{equation}{section}
\numberwithin{figure}{section}

\section[\appendixname~\thesection]{Cosmological equations for \texorpdfstring{\texttt{class}}{}}\label{app:B}
\texttt{class} uses the number of e-folds ($N \equiv \text{ln}\, a$) to evolve the equations, but at any time one can also make use of the cosmic time ($t$) or conformal time ($\eta$). Since our CHDE model is given in terms of the conformal time, we write first order cosmological equations as (with units $\frac{8 \pi G}{3} =1$):
\begin{eqnarray} \label{eqn:class}
\frac{d \rho_{dm}}{dN} &=& -3  \rho_{dm} - \frac{n Q}{3a} \frac{H_0^3}{ H} \, (H_0 \eta)^{n-1} ,  \\
\frac{d t}{dN} &=& \frac{1}{H} , \\
\frac{d \eta}{dN} &=& \frac{1}{a\, H} , \\
\frac{d r_{s}}{dN} &=& \frac{c_s}{ a H} , \\
\frac{d D}{dN}
&=&
- \frac{1}{2} \frac{d h}{dN} + \frac{1}{\rho_{dm}} \frac{d\rho_{\cal M}}{dN} \, D ,   \label{eqn:classB5}
\end{eqnarray}
whereby we employed the synchronous gauge with $\theta_{dm} =0$ and $\delta \rho_{\cal M} =0$. The Hubble parameter is:

\begin{equation}\label{eqn:H2}
    H^2 = \frac{\rho_b^{(0)}}{a^3} + \frac{\rho_\gamma^{(0)}}{a^4}
    + \frac{k}{a^2} +
    \rho_{dm} +
          \frac{Q H_{0}^2}{3} (H_{0} \eta)^n
\end{equation}
and the following definitions apply:
\begin{eqnarray}\label{eqn:defs}
a d\eta   &\equiv&  dt , \, \text{cosmic time}  \\
c_{s}   &\equiv&  \frac{1}{\sqrt{3 \left( 1 + \frac{3 \rho_b}{4 \rho_\gamma}\right)}}, \, \text{plasma sound speed},  \\
r_{s} &\equiv& \int_0^{\eta} d \eta^{\prime} \, c_{s}(\eta^{\prime}),  \, \text{sound horizon}, \\
D   &\equiv&  \delta_{dm} , \, \text{matter density contrast}.
 \end{eqnarray}

\end{document}